\documentstyle[11pt,paspconf]{article}

\begin{document}

\title{Resolving the C IV and Mg II Absorption in NGC~4151}

\author{Gerard A.\ Kriss}
\affil{Department of Physics and Astronomy, The Johns Hopkins University,
    Baltimore, MD 21218}

\begin{abstract}
Observations of the {\sc C~iv} and Mg~{\sc ii} absorption lines in the
Seyfert 1 galaxy NGC~4151 obtained with the GHRS in October 1994 are presented.
The data from the STScI archive show multiple broad and narrow
components in both species.  In addition to Galactic absorption, four narrow
and four broad systems associated with NGC~4151 are identified.
Two broad systems dominate the total equivalent width, and their
mean blueshift and width are comparable to the broad Lyman line
and continuum absorption seen in far-UV spectra from the Hopkins
Ultraviolet Telescope.
Narrow-line {\sc C~iv} emission is present on the red side of the broadest
absorption trough, and narrow absorption at the systemic velocity of NGC~4151,
presumably in its own ISM, absorbs the core of the narrow emission line.
Strong Mg~{\sc ii} absorption is present in all but two velocity systems.
Ratios relative to the corresponding {\sc C~iv} components suggest a
low ionization parameter for the absorbing gas:
$U \sim 1-3 \times 10^{-3}$.
This makes none of the identified UV absorption systems
a good candidate for association with the warm X-ray absorbing gas.
\end{abstract}


\section{UV and X-ray Absorption in NGC 4151}

A persistent problem in understanding the absorbing material in NGC~4151 has
been reconciling the vastly different gas columns inferred for the X-ray
absorption and for the UV absorption.
The X-ray absorbing column varies between $10^{22}$ and $10^{23}~\rm cm^{-2}$.
Bromage et al. (1985) estimated a total column for the UV-absorbing material
of no more than $\sim 10^{21}~\rm cm^{-2}$.
The neutral hydrogen column is variable (Kriss et al. 1995).
The bulk of the absorption is in low column density gas with
$\rm N_H \sim 10^{18}~\rm cm^{-2}$ and Doppler parameter
$\rm b \sim 300~km~s^{-1}$.
Any low-b component has a neutral column no greater than
$5 \times 10^{20}~\rm cm^{-2}$.

\begin{figure}[t]
\plotfiddle{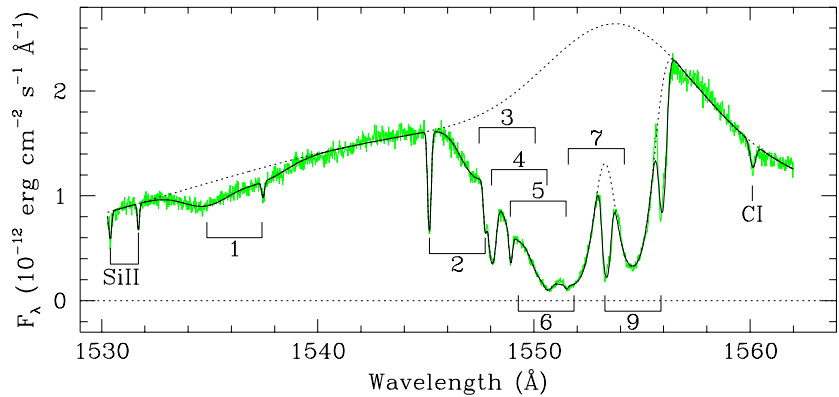}{1.85in}{-90}{50}{50}{-210}{158}
\caption{
The {\sc C~iv} profile observed with the GHRS is shown as a gray line.
The smooth solid line is the best fit to the modeled emission and
absorption profile.
The smooth dotted line shows the intrinsic shape of the modeled broad
emission plus continuum; another dotted line shows the intrinsic profile
of narrow-line emission on the red side of the absorption trough.
Eight {\sc C~iv} absorption doublets are indicated as well as absorption
by Si~{\sc ii} and Galactic {\sc C~i}.
}
\end{figure}

One possibility for reconciling these differences has been the recent success
of warm absorber models for characterizing the X-ray absorption and the
associated UV absorption lines in 3C~351 and NGC~5548
(Mathur et al. 1994; Mathur et al. 1995).
In such models the absorption arises in gas
photoionized by the central engine (e.g., Netzer 1993;
Krolik \& Kriss 1995).
The X-ray absorption is dominated by highly ionized species of heavy ions
(e.g., {\sc O~vii} and {\sc O~viii}).
The total gas columns can be quite high ($10^{22}$--$10^{23}~\rm cm^{-2}$),
with relatively low columns in the lower ionization species responsible
for the UV absorption.
Warm absorber models with a reflection component
can fit the X-ray spectrum of NGC~4151 (Weaver et al. 1994a,b).
Kriss et al. (1995) find that similar models can also account for the
high ionization lines in NGC~4151 (e.g., {\sc O~vi}, {\sc N~v}, and {\sc
C~iv}),
but they cannot simultaneously match the particularly strong absorption
in lower ionization species such as {\sc H~i}, {\sc C~iii}, and Si~{\sc iv}.
They conclude that a single-zone warm absorber is insufficient.
To search for absorption components that might possibly be identified
with the X-ray absorbing gas, I examined archival high resolution GHRS
spectra of the {\sc C~iv} and Mg~{\sc ii} line profiles in NGC~4151.

\section{Observations and Analysis}

Fig.\,1 shows the spectrum of NGC~4151 in the {\sc C~iv} region
with 14 $\rm km~s^{-1}$ resolution obtained in 8486 s
using grating G160M of the GHRS on 28 October 1994.
A model consisting of an underlying power law continuum, three
broad Gaussian emission lines, and
8 {\sc C~iv} absorption line doublets fits the data well and gives
$\chi^2 = 1998$ for 1800 points and 50 free parameters.
Although the deepest and broadest {\sc C~iv} doublet is saturated, the bottom
of the line profile is not black.  Either this gas only partially covers the
source (at the 90\% level, both continuum and broad line),
or 10\% of the continuum flux is scattered around the absorbing region back
into our line of sight.
Narrow-line emission is visible on the red side of the
{\sc C~iv} absorption trough.
This emission is apparently unabsorbed by the broad absorbing gas;
a final layer of absorbing gas, however, lying at the systemic velocity
of NGC~4151, absorbs the core of the narrow-line profile.
This is presumably the ISM or halo of NGC~4151.

The spectrum of the Mg~{\sc ii} region at 10 $\rm km~s^{-1}$ resolution
obtained in a 1414 s integration
with grating G270M of the GHRS on 29 October 1994 is shown in Fig.\,2.
The best fit to the modeled emission and absorption profile gives
$\chi^2 = 1636$ for 1438 points and 22 free parameters.
As with {\sc C~iv}, the Mg~{\sc ii} emission was modeled with 3 Gaussians.
Seven Mg~{\sc ii} absorption doublets are required.
Table 1 gives the velocities, equivalent widths, Doppler
parameters, and column densities of each of the absorption components
fit in the {\sc C~iv} and the Mg~{\sc ii} spectra.

\begin{figure}[t]
\plotfiddle{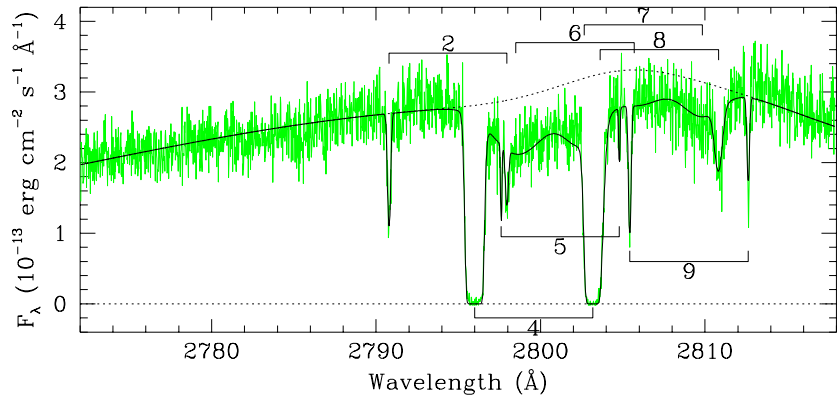}{1.85in}{-90}{50}{50}{-210}{158}
\caption{
The Mg~{\sc ii} profile observed with the GHRS is
shown as the gray line.
The smooth solid line is the best fit to the modeled emission and absorption
profile; the smooth dotted line is the modeled broad
emission plus underlying continnum.
Seven Mg~{\sc ii} absorption doublets are marked.
Components 1 and 3 from the {\sc C~iv} profile have no Mg~{\sc ii}
counterparts;
component 9 in the Mg~{\sc ii} profile has no {\sc C~iv} counterpart.
The deepest absorption doublet, component 4, is Galactic Mg~{\sc ii}.
}
\end{figure}

\begin{center}
\renewcommand{\tabcolsep}{4pt}
\small
\begin{tabular}{| c | c c c c || c c c c |}
\multicolumn{9}{c}{\normalsize Table 1.  Absorption Line Components in
NGC~4151} \\[2pt]
\hline
&&&&&&&&\\[-8pt]
 & \multicolumn{4}{c||}{\sc C~iv} & \multicolumn{4}{c|}{Mg~\sc ii} \\[4pt]
\hline
&&&&&&&&\\[-8pt]
\# &  $cz_{\odot}$  & EW   & {\it b} & $\rm N_{CIV}$ &  $cz_{\odot}$ & EW   &
{\it b} & $\rm N_{MgII}$ \\
     & $\rm ( km~s^{-1} )$ & (\AA) & $\rm ( km~s^{-1} )$ & $\rm ( cm^{-2} )$ &
$\rm ( km~s^{-1} )$ & (\AA) & $\rm ( km~s^{-1} )$ & $\rm ( cm^{-2} )$ \\[4pt]
\hline
&&&&&&&&\\[-8pt]
1 & $-2561$          &  0.514 & 294           & $1.4\times10^{14}$ & \ldots &
\ldots & \ldots & \ldots \\
2 & $\phantom{0}$$-567$ &  0.120 & $\phantom{0}20$ & $4.0\times10^{13}$ &
$-577$
 & 0.143 & $\phantom{0}13$ & $4.6\times10^{12}$ \\
3 & $\phantom{0}$$-128$ &  0.642 & 203           & $1.8\times10^{14}$ & \ldots
&
 \ldots & \ldots & \ldots \\
4 & $\phantom{000}$$-4$ &  0.310 & $\phantom{0}43$ & $1.1\times10^{14}$ &
$\phantom{0}$$-19$ & 1.259  & $\phantom{0}34$ & $4.1\times10^{14}$ \\
5 & $\phantom{-0}162$  &  0.083 & $\phantom{0}19$ & $2.5\times10^{13}$ &
$\phantom{-}154$ & 0.052 & $\phantom{00}6$ & $1.5\times10^{12}$ \\
6 & $\phantom{-0}187$  &  1.026 & 163           & $3.4\times10^{14}$ &
$\phantom{-}276$ & 1.116 & 235 & $2.9\times10^{13}$ \\
7 & $\phantom{-0}671$  &  4.018 & 234           & $5.3\times10^{15}$ &
$\phantom{-}685$ & 0.852 & 176 & $2.3\times10^{13}$ \\
8 & \ldots & \ldots & \ldots & \ldots & $\phantom{-}799$ & 0.329  &
$\phantom{0}
33$ & $1.0\times10^{13}$ \\
9 & $\phantom{-}1020$   &  0.407 & $\phantom{0}43$ & $1.7\times10^{14}$ &
$\phantom{-}992$ & 0.134 & $\phantom{0}11$ & $4.4\times10^{12}$ \\[4pt]
\hline
\end{tabular}
\end{center}

\section{Photoionization Models of the Absorbing Gas}

For the absorption components intrinsic to NGC~4151,
I assume that the gas is photoionized by the active nucleus.
Computing photoionization models similar to those discussed by Krolik \&
Kriss (1995) and Kriss et al. (1996), I search for ionization
parameters and total column densities
that match the Mg~{\sc ii} and {\sc C~iv} columns seen in the data.
Table 2 summarizes the column density ratios of each of the
absorption components and the matching ionization parameters and total
column densities.  The velocities are now relative to the systemic velocity
of NGC~4151 ($993~\rm km~s^{-1}$, Mundell et al. 1995).

\begin{center}
\begin{tabular}{| c | c | c | c | c |}
\multicolumn{5}{c}{Table 2.  Photoionization Models of the Components} \\[2pt]
\hline
&&&&\\[-10pt]
\# & $\Delta v$  & $\rm N_{MgII} / N_{CIV}$ & log {\it U} & log $N_{total}$ \\
   & $\rm ( km~s^{-1} )$ &                      &     & $\rm (cm^{-3})$
\\[4pt]
\hline
&&&&\\[-10pt]
1 & $-3553$             & $< 0.02$         &  $> -2.6$ & \ldots \\
2 & $-1559$             &  0.12\phantom{0} &  $-2.9$   & 18.3 \\
3 & $-1120$             & $< 0.02$         &  $> -2.6$ & \ldots \\
4 & $\phantom{0}$$-$992 & 3.73\phantom{0}  &  {Galactic} & 20.3 \\
5 & $\phantom{0}$$-$830 & 0.060            &  $-2.7$   & 18.1 \\
6 & $\phantom{0}$$-$805 & 0.085            &  $-2.8$   & 18.2 \\
7 & $\phantom{0}$$-$321 & 0.004            &  $-2.1$   & 19.9 \\
8 & $\phantom{0}$$-$193 & $> 5.0$          &  $< -3.4$ & 17.0--18.0 \\
9 & $\phantom{000}$$-$1   & 0.026            &  $-2.6$   & 18.6 \\[4pt]
\hline
\end{tabular}
\end{center}

Note that all the absorbing systems have fairly low ionization parameters.
None of the systems in which Mg~{\sc ii} absorption is visible is a good
candidate for association with the warm X-ray absorbing gas, which typically
has high ionization parameters $U \sim 1$
and high total column densities log $N_{total} \sim 23$ (Weaver et al.
1994a,b).
While components 1 and 3 might be possible candidates, note that component 1
is visible only at this single epoch.  It is absent from all other GHRS
observations at both higher and lower flux levels (Weymann et al. 1997).
Observations of higher
ionization species such as Si~{\sc iv} or {\sc N~v} are required to
set more stringent constraints on the ionization parameters and the total
column densities of components 1 and 3.

\acknowledgments

This work was supported by NASA LTSA grant\\
NAGW-4443 to the Johns Hopkins University.

\end{document}